\documentclass{iau} 

\usepackage{graphicx,caption,subcaption}
\usepackage{aas_macros}

\title[NFFF extrapolation of AR12192] %% give here short title %%
{Study of magnetic field topology of active region 12192 using an extrapolated non-force-free magnetic field}

\author[Prasad et al.]   %% give here short author list %%
{A. Prasad$^1$, R. Bhattacharyya$^1$, Q. Hu$^2$, S. S. Nayak$^1$ \& Sanjay Kumar$^3$}

\affiliation{$^1$ Udaipur Solar Observatory, Physical Research Laboratory\\
$^2$Center for Space Plasma and Aeronomic Research, The University of Alabama in
Huntsville\\
$^3$Post Graduate Department of Physics, Patna University}

\pubyear{2018}
\volume{340}  %% insert here IAU Symposium No.
\jname{Long – Term Datasets for the Understanding of Solar and Stellar Magnetic Cycles}
\editors{D. Banerjee, J. Jiang, K. Kusano \& S. Solanki}
\begin{document}

\maketitle

\begin{abstract}
The solar active region (AR) 12192 was one of the most flare productive region of solar cycle 24, which produced many X-class flares; the most energetic being an X3.1 flare on October 24, 2014 at 21:10 UT. Customarily, such events are believed to be triggered by magnetic reconnection in coronal magnetic fields. Here we use the vector magnetograms from solar photosphere, obtained from Heliospheric Magnetic Imager (HMI) to investigate the magnetic field topology prior to the X3.1 event, and ascertain the conditions that might have caused the flare. To infer the coronal magnetic field, a novel non-force-free field (NFFF) extrapolation technique of the photospheric field is used, which suitably mimics the Lorentz forces present in the photospheric plasma. We also highlight the presence of magnetic null points and quasi-separatrix layers (QSLs) in the magnetic field topology, which are preferred sites for magnetic reconnections and discuss the probable reconnection scenarios.

\end{abstract}

\firstsection % if your document starts with a section,
\section{Introduction}
The solar corona represents a magnetized plasma with high electrical conductivity whose evolution is determined by magnetohydrodynamics (MHD). The large magnetic Reynolds number $R_M (= vL/\eta$, in usual notations) of the corona ensures that the magnetic field lines (MFLs) to remain tied to fluid parcels during their evolution. In contrast, instances of various eruptive events (flares and coronal mass ejections) occurring in the corona are believed to be signatures of magnetic reconnection: the topological rearrangement of magnetic field lines along with the conversion of magnetic energy into heat and the kinetic energy of mass  motion. Currently, the coronal fields are numerically extrapolated from the photospheric magnetic fields, as direct coronal measurements are not available. Here we present a model for coronal magnetic fields derived from the variational principle of the minimum energy dissipation rate (\cite[Hu et al. 2008]{2008ApJ...679..848H}; \cite[Prasad et al. 2017]{2017ApJ...840...37P}) formulated by superposing three linear-force-free fields and its application for the X3.1 flare event of AR 12192 on October 24th 2014 at 21:10 UT.

\section{Results}
\begin{figure}[hp]
  \centering
  \begin{subfigure}[]{0.45\textwidth}
    \centering
    \includegraphics[width=1\linewidth]{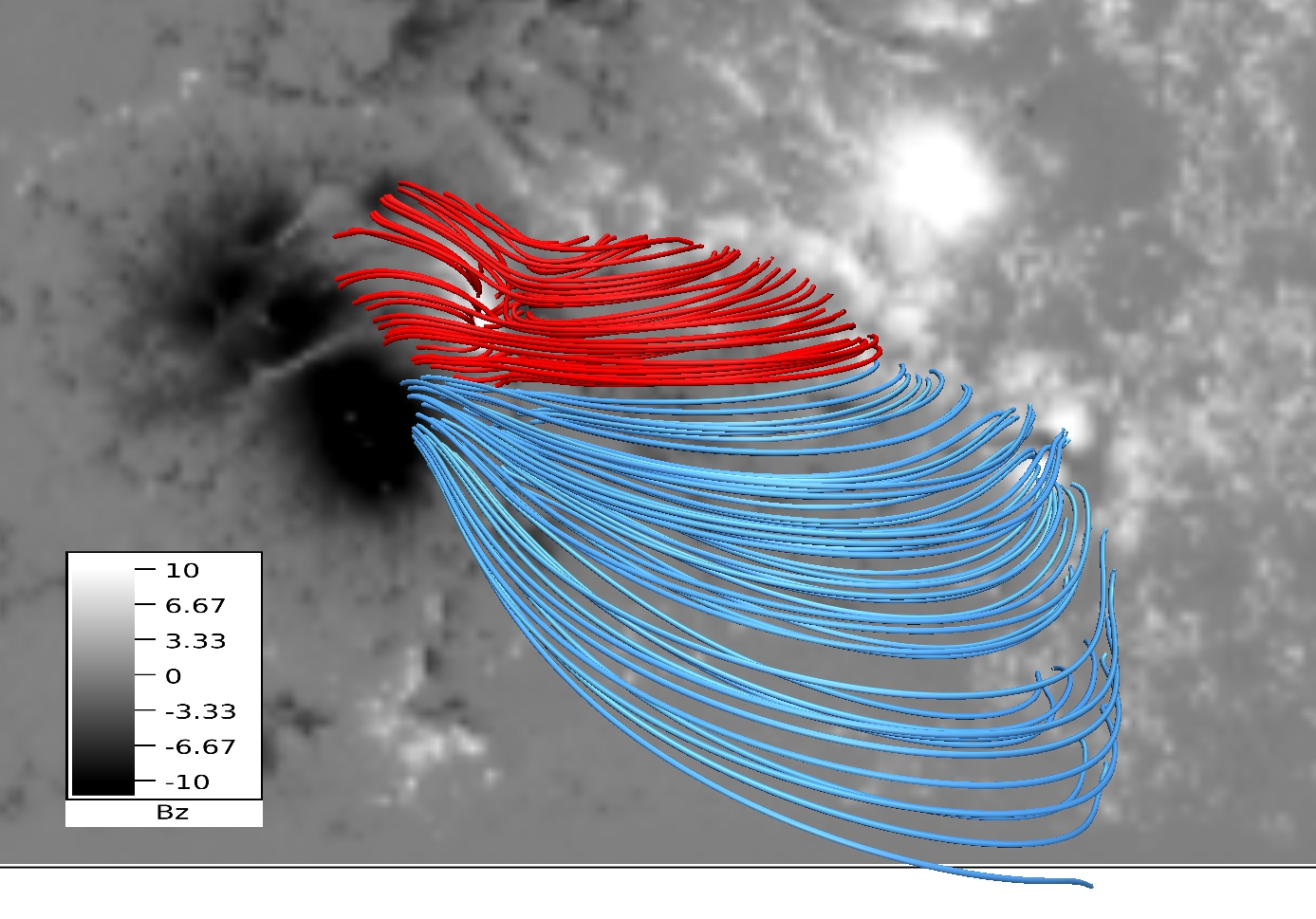}
    \caption{}
    \label{fig1a}
  \end{subfigure}
\quad
  \begin{subfigure}[]{0.45\textwidth}
    \centering
    \includegraphics[width=1\linewidth]{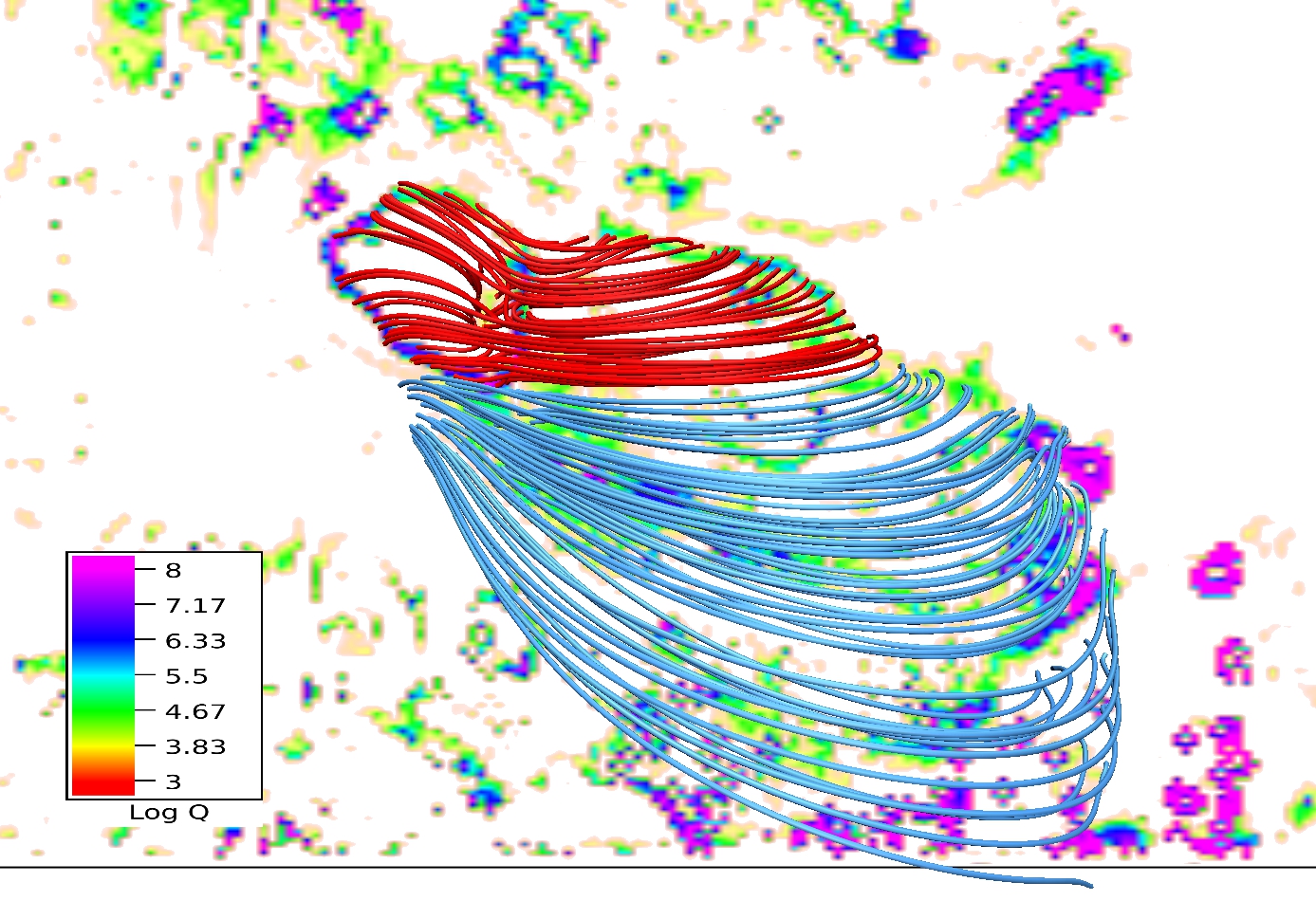}
    \caption{}
    \label{fig1b}
  \end{subfigure}
   \caption{The extrapolated MFLs for AR 12192 overlaid with (a) vertical component of magnetic field $Bz$ and (b) log of the squashing-factor $Q$. The values for $B_z$ are scaled by a factor of 200.}
  \label{fig1}
\end{figure}

For the NFFF extrapolation, we select the vector magnetogram at 20:46 UT obtained from HMI on board the Solar Dynamics Observatory (SDO), roughly 20 minutes prior to the flare. In Fig. \ref{fig1a}, we show the large-scale sheared MFLs (shown in blue) near the polarity-inversion line and also the MFLs constituting the spine-fan structure of a three-dimensional (3D) null point (shown in red). Importantly, these MFLs  demarcate the QSLs (regions of high squashing-factor $Q$; c.f. Fig. \ref{fig1b}). It is well known that 3D nulls and QSLs are preferred sites for magnetic reconnections \cite[(Pontin 2012)]{2012RSPTA.370.3169P}. 

\begin{figure}[hp]
  \centering
  \begin{subfigure}[]{0.45\textwidth}
    \centering
    \includegraphics[width=1\linewidth]{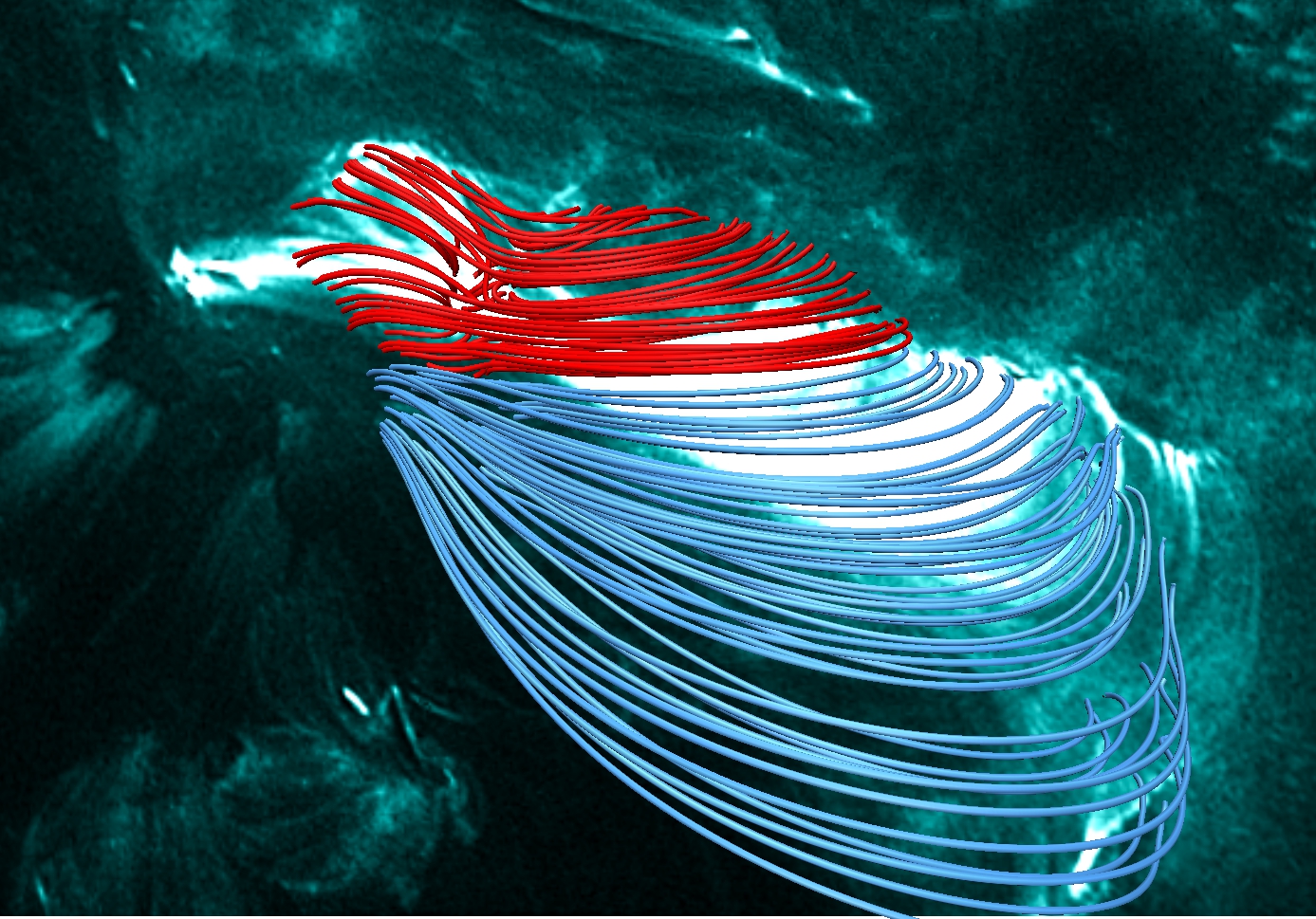}
    \caption{}
    \label{fig2a}
  \end{subfigure}
\quad
  \begin{subfigure}[]{0.45\textwidth}
    \centering
    \includegraphics[width=1\linewidth]{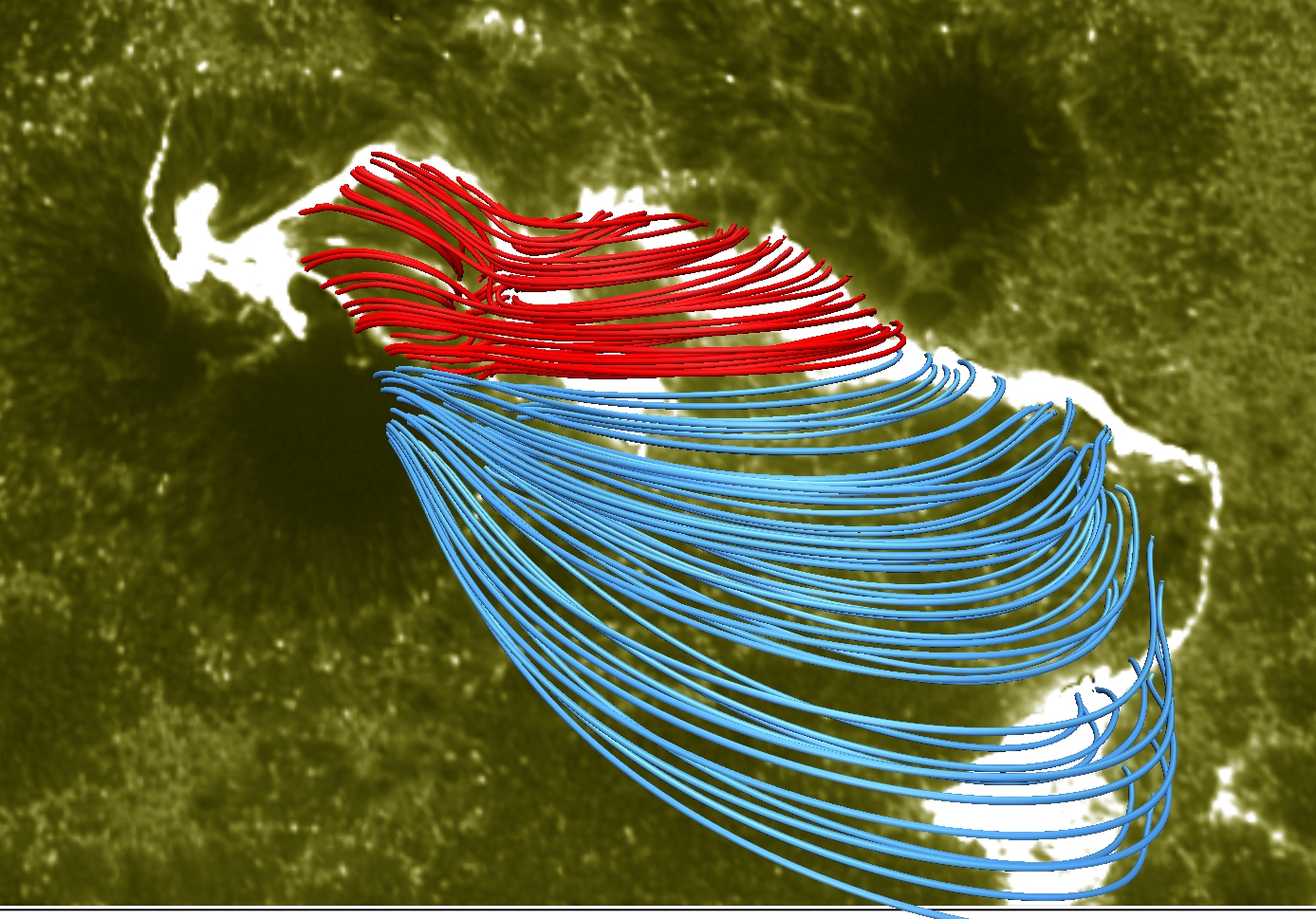}
    \caption{}
    \label{fig2b}
  \end{subfigure}
   \caption{The extrapolated MFLs for AR 12192 overlaid with images of (a)  AIA 131 \AA~ at 21:02 UT and (b) AIA 1600 \AA~ at 21:28 UT. }
  \label{fig2}
\end{figure}

In Fig. \ref{fig2}, we overlay the above MFLs with the extreme-ultra-violet (EUV) 131 \AA~ channel and the ultra-violet (UV) 1600 \AA~ channel images from the Atmospheric Imaging Assembly (AIA) at 21:02 UT and 21:28 UT respectively, when the brightening in these channels are maximum. The near accurate correspondence between the brightening observed in the AIA channels with the MFL footpoints validate the NFFF extrapolations presented here. The dynamics of the MFLs leading to the flare require full MHD simulations with these NFFF as initial fields. This would be presented elsewhere in future.

\end{document}